\documentclass{article}
\usepackage{spconf,amsmath,graphicx}
\usepackage[bb=boondox]{mathalfa}

\newcommand{\matr}[1]{\mathbf{#1}} 

\title{MUSIC REARRANGEMENT USING HIERARCHICAL SEGMENTATION}
\name{Christos Plachouras, Marius Miron}
\address{Music Technology Group, Universitat Pompeu Fabra, Spain}
\begin{document}
\ninept
\maketitle
\begin{abstract}
Music rearrangement involves reshuffling, deleting, and repeating sections of a music piece with the goal of producing a standalone version that has a different duration. It is a creative and time-consuming task commonly performed by an expert music engineer. In this paper, we propose a method for automatically rearranging music recordings that takes into account the hierarchical structure of the recording. Previous approaches focus solely on identifying cut-points in the audio that could result in smooth transitions. We instead utilize deep audio representations to hierarchically segment the piece and define a cut-point search subject to the boundaries and musical functions of the segments. We score suitable entry- and exit-point pairs based on their similarity and the segments they belong to, and define an optimal path search. Experimental results demonstrate the selected cut-points are most commonly imperceptible by listeners and result in more consistent musical development with less distracting repetitions.

\end{abstract}
\section{Introduction}\label{sec:intro}

Altering the duration of music pieces is an important part of audiovisual content creation for advertisements, documentaries, film, short-form videos, vlogs.
Rearranging music is often the preferred approach for music retiming, because the alternatives come with inherent disadvantages: time-stretching may reduce audio quality and alter the intended feeling of the composition, while fades may remove the intended start and end of a piece, resulting in a more abrupt or unpolished experience. Music rearrangement is, however, a tedious process often delegated to expert music engineers, who have to spend time listening to the recording, understanding its structure, and performing suitable edits.

The democratization of video editing has brought even more interest in music rearrangement, as reflected by the rearrangement systems video editing software are starting to integrate~\cite{adoberemix}. In spite of its commercial appeal, the task of automatic music rearrangement has only sparingly been explored in scientific research 
and has appeared with a variety of names used almost synonymously, including music retargeting~\cite{retargeting, genetic}, resynthesis~\cite{dancefloor}, retiming~\cite{adoberemix}, and remixing~\cite{adoberemix}. 

To avoid ambiguity, we define a rearrangement of the recording of a piece of music to be a standalone piece that has a different duration to the original and that is constructed solely from segments of the original piece. It should have the same beginning and ending as the original and have smooth transitions between the reshuffled segments without unnatural discontinuities of music information such as melodies, chords, and instrumentation. In contrast to music summarization~\cite{logan2000music} and thumbnailing~\cite{bartsch2005audio} the rearrangement needs to stand as a music piece on its own. It is designed to be experienced by listeners, rather than be used as a proxy representation for other tasks such as music classification~\cite{raposo2016using}. Furthermore, unlike in music remixing~\cite{yang2020remixing}, a rearranged piece must solely include unedited segments from the original piece.

Our primary contribution in this work is the introduction of a novel automatic music rearrangement approach leveraging the hierarchical structure of the music piece. Unlike previous approaches which frame rearrangement as a suitable cut-point identification task and do not have inherent considerations for musical development (see Sec. \ref{sec:related_work}), we anchor rearrangement and cut-point identification to the extracted segment boundaries and musical functions. Additionally, we introduce the use of deep audio features from a music auto-tagging model to estimate the perceptibility of segment transitions. Finally, we provide a freely accessible, open-source, modular implementation of our method \cite{webpage}.

\section{Related work}\label{sec:related_work}
Previous approaches frame music rearrangement as the task of identifying suitable cut-points, pairs of entry- and exit-frames in the audio between which a jump can be performed without any perceived discontinuity~\cite{constrained,genetic,dancefloor, retargeting,loudness}. While smooth transitions are critical for creating a consistent rearrangement, we argue that this approach does not give adequate consideration to the musical development of the rearrangement. A segment of the recording might get selected to succeed another because it shares a small snippet of audio with it, but there is no guarantee that it will feel like the natural continuation of the current segment, nor that it contributes to a consistent development of musical ideas in the rearrangement. This means that inevitably these systems~\cite{constrained,genetic,dancefloor} can lead to unnatural, distracting repetitions \cite{retargeting}, or inconsistent, rushed, or dull evolution of musical ideas given the rearrangement's duration.

Wenner et al.~\cite{retargeting} partially try to address this issue by analyzing the music piece's structure and ensuring jump-points don't occur between segments of the same type. While this approach alleviates some of the potentially unnatural repetitions, it does not fully address musical development in the rearrangement. Unlike our approach where segments are automatically reshuffled, the authors instead present some ideas for manually editing the resulting structure of the rearrangement.

Towards improving the imperceptibility of transitions, a Convolutional Neural Network was used to classify whether a frame contains a good transition or cut-point ~\cite{loudness}.

We instead opt to use embeddings from a model that has been trained on a variety of styles~\cite{musictag}, is robust to recording conditions~\cite{musictag}, and has been shown to improve hierarchical music segmentation~\cite{deepemb}. We use these embeddings both for analyzing the hierarchical structure of pieces, and for identifying smooth transitions across and within segments.

To test the relevance of musical development and the suitability of the deep embeddings, we run a listening test (see Sec. \ref{sec:evaluation}) on rearrangements of five songs from different music traditions and recording conditions. In contrast to previous approaches who only evaluate transition quality, we evaluate development (e.g. no abrupt changes in musical ideas) and balance (e.g. lack of repetitions), along with consistency (e.g. good cut-points).

Conceptually, our work is also related to that of Thalmann et al.~\cite{thalmann2019representing}, where dynamic music objects existing at multiple hierarchies are used as modifiable and reusable music content for the web. To that extent, our task is defined by constraints regarding the input modality (audio), the fixed duration, and the properties of the resulting piece (good development, consistency, and balance).  


\section{Methods}\label{sec:methods}
Our proposed method has the following steps: (1) we construct an encoding capturing the global patterns formed by various musical elements as well as their evolution over time, (2) we decompose the encoding hierarchically to uncover groupings of musical ideas at various scales, (3) we identify and rank suitable transition points across and within structural segments, and (4) we define rearrangement as an optimal path search using the extracted transition points.

\subsection{Structure encoding}\label{subsec:structure_encoding}
When rearranging a music recording, a sound engineer inevitably changes its structure by removing, repeating, or reshuffling musical ideas such as melodic phrases, chord progressions, lyrics, and others. Towards avoiding their interruption, important care is given to choosing cut-points that give good transitions between groupings of these ideas~\cite{thalmann2019representing}. These groupings exist at various temporal scales and may have hierarchical relationships between them; for example, a group of elements forming the chorus of a song can have a repeated chord sequence, which in turn can have melodic phrases each of which can be contained in a measure. We will try to uncover these hierarchical groupings using the hierarchical structure analysis method proposed by McFee and Ellis~\cite{spectral}, with the subsequent enhancements proposed by Salamon et al. \cite{deepemb}.

We first compute the beat times $\matr{b} = \{b_1\ ..\ b_M\}$ and downbeat times $\matr{o} = \{o_1\ ..\ o_M\}$ from the audio, with $M$ and $N$ the total number of beats and downbeats respectively. To do this, we replace the beat tracking used by the previous works with BeatNet \cite{beatnet}, which uses a Recurrent Convolutional Neural Network and particle filtering to improve beat tracking performance and also provide downbeat and meter tracking. We aim to create an encoding that captures global patterns in various music elements such as harmony, instrumentation, mood, and energy, but also the homogeneity of successive frames. To do this we use the features proposed by Salamon et al \cite{deepemb}: deep embeddings $\matr{X^{T\times E}}$ learned from a music auto-tagging model~\cite{musictag} and the Constant-Q Transform $\matr{Y^{T\times F}}$ of the audio as repetition features, and deep embeddings $\matr{Z^{T\times H}}$ from a few-shot sound event detection model~\cite{fewshot} to encode homogeneity, where $T$ refers to total number of time frames and $E,F,H$ to the respective feature dimensions. We beat-synchronize all features to the beat track $\matr{b}$ by aggregating the feature vectors of frames $T$ belonging to each beat, and we obtain the corresponding beat-synchronized feature matrices $\matr{\hat{X}^{N\times E}},\matr{\hat{Y}^{N\times F}},\matr{\hat{Z}^{N\times H}}$.

\begin{figure}[]
    \centering
  \includegraphics[width=0.45\textwidth]{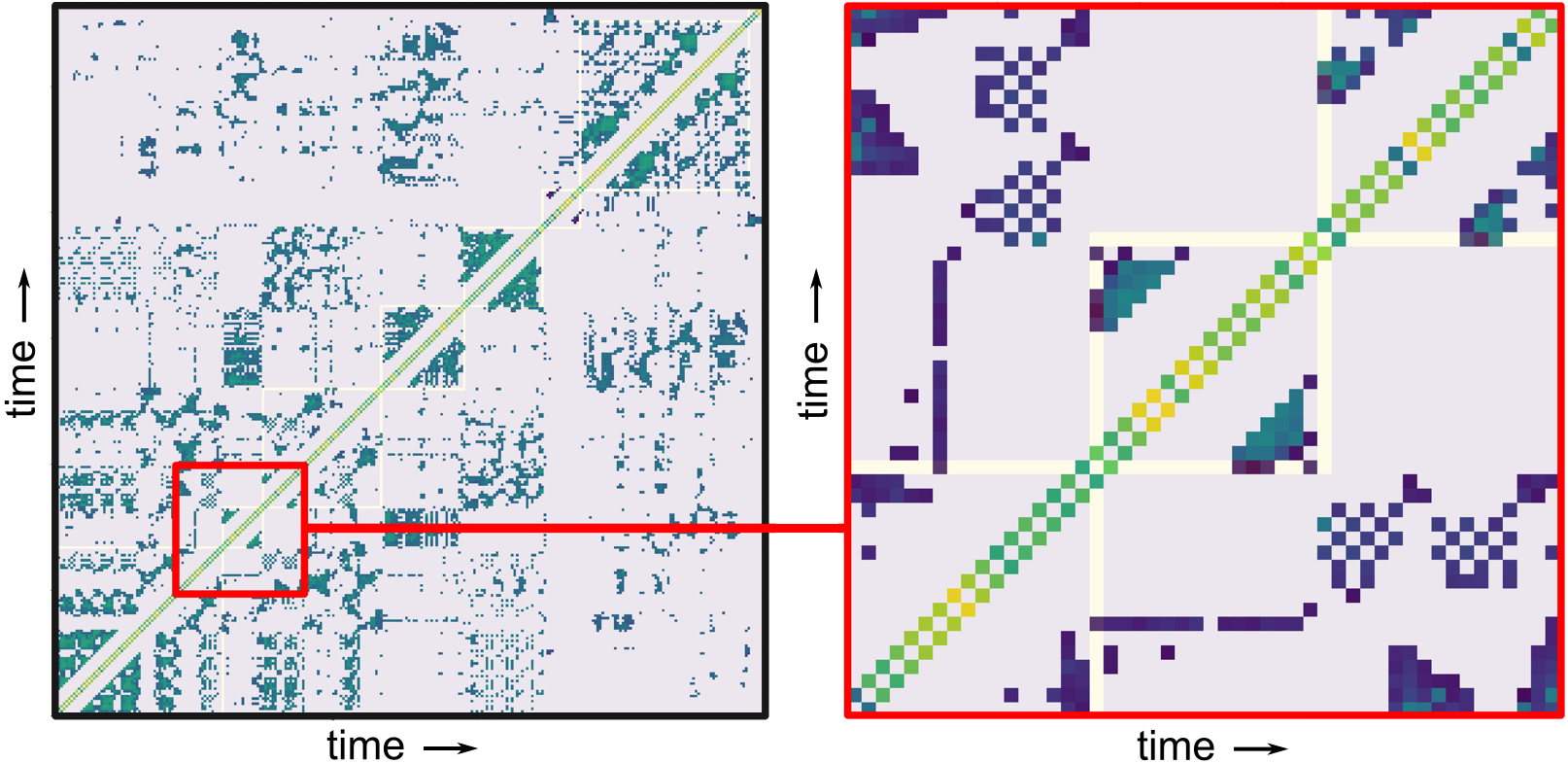}
  \caption{\label{figure}Recurrence matrix $\matr{R}$ with segment overlay in light yellow}
  \label{fig:rec}
\end{figure}

As proposed by McFee and Ellis \cite{spectral}, we compute weighted, undirected recurrence graphs from the repetition features such that
\begin{equation}
\matr{R}_\matr{X}(i,j) =
\left\{
\begin{array}{ll}
      exp({-\sqrt{|\matr{\hat{x}}_i-\matr{\hat{x}}_j|^2}/\mu}) & \matr{\hat{x}}_i, \matr{\hat{x}}_j \text{ mutual kNN} \\
      0 & \text{otherwise}
\end{array},
\right. \label{eq:recm1}
\end{equation}
where $\matr{\hat{x}}_i$ represents the $E$-dimensional column of $\matr{\hat{X}}^{N\times E}$ at beat $\matr{b}_i$ and $\mu$ represents the median distance between furthest nearest neighbors. The recurrence matrix $\matr{R}_\matr{Y}$ is computed in the same way from $\matr{\hat{Y}}^{N\times F}$. Importantly, since similar beats are connected, diagonals in these matrices are consecutive connected beats that indicate a repeated pattern (see Fig. \ref{fig:rec}).

To encode homogeneity, we construct a sequence matrix from $\matr{\hat{Z}}^{N\times H}$ of the distances of each beat with its immediate neighbors such that
\begin{equation}
\matr{R}_\matr{Z}(i,j) = 
\left\{
\begin{array}{ll}
    exp({-\lvert \matr{\hat{Z}}_i - \matr{\hat{Z}}_j \lvert ^2/\sigma^2}) & |i-j|=1 \\
    0 & \text{otherwise}
\end{array},
\right. \label{eq:recm2}
\end{equation}
where $\sigma$ is the median distance between beats.

We then compute a weighted sum of the matrices as such:
\begin{equation}
\matr{R} = 0.25 \matr{R}_{\matr{X}} + 0.25 \matr{R}_{\matr{Y}} + 0.5 \matr{R}_{\matr{Z}} 
\label{eq:recm3}
\end{equation}

We refer to Salamon et al.~\cite{deepemb} for further details of their implementation, combination weights, and other parameter values that resulted in an improvement in hierarchical structure analysis.

\subsection{Hierarchical segmentation}\label{subsec:hierarchical_segmentation}
\begin{figure}[htbp!]
    \centering
  \includegraphics[width=0.48\textwidth]{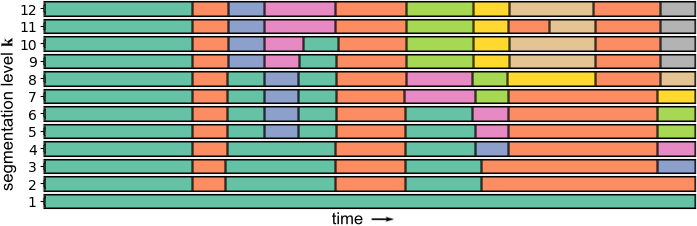}
  \caption{\label{figure}Hierarchical structure with $k =\{1\ ..\ 12\}$ levels}
  \label{fig:structure}
\end{figure}

As proposed in the related works \cite{spectral, deepemb}, we first compute the symmetric normalized Laplacian $\matr{L}$ of $\matr{R}$:
\begin{equation}
\matr{L} = \matr{I}-\matr{D}^{-1/2}\matr{R}\matr{D}^{-1/2},
\label{eq:decomposition}
\end{equation}
where $\matr{D}$ is the diagonal matrix of $\matr{R}$ and $\matr{I}$ is the identity matrix. 
We define $\mathcal{E}_k$ as the set containing the first $k$ eigenvectors of $\matr{L}$, where $k=\{1\ ..\ 12\}$. Sets with a higher $k$ will contain a more granular representation of $\matr{L}$. We do spectral clustering \cite{spectraltutorial} on each set $\mathcal{E}_k$ using $k$ clusters, therefore increasing the granularity of analysis as the representation granularity increases. The result is a multi-level segmentation, where in each level every beat is assigned a segment type, and the cluster change-points determine the segment boundaries (see Fig. \ref{fig:structure}). We construct the global set of segments $\mathcal{S}$ across all levels, where each of the generated segments is defined by its level $k$ and a pair of beat indices $\{p, q\}$ denoting its starting and ending boundary.

While the beat-level analysis was important for making granular comparisons, musically the measure (delimited by two consecutive downbeats) is too fundamental to interrupt. We therefore opt to quantize the beat-level segmentation by replacing boundary beats in $\mathcal{S}$ with their closest downbeat from the downbeat track $\matr{o}$ we computed. For a music piece with 120 beats per minute (BPM), quantization limits the duration precision of the rearrangement to $\pm2$ seconds. Precision also depends on the duration of the shortest segment extracted from the segmentation; if the user requires predicable precision within a certain threshold, they might set the segmentation to stop when for a level $k$ a segment of a certain duration is produced, rather than using a fixed number of levels.

\subsection{Transition identification}\label{seg:cut-point}
We categorize our suitable transition search into two types: segment transition points, which refer to transitions between structural segments, and internal transition points, which refer to transitions that happen within a single structural segment.

\subsubsection{Segment transition points}\label{subsubsec:segment_transition_points}
We conceptualize rearrangement as the task of selecting and ordering segments from $S$ so that the sum of their durations in seconds is close to the target duration for the rearrangement.
One key problem of previous approaches is the resulting distracting repetitions~\cite{retargeting}. To avoid these, we take advantage of the segment labels assigned from spectral clustering. Segments of the same type will have similar music information, and as the cluster number and hence segmentation level $k$ increases, segments with the same label will likely have even stronger similarity. When considering segments to succeed the current segment, we will therefore eliminate segments with the same label.

While the segmentation helps avoid interrupting groupings of ideas, this does not guarantee that jumps between segment boundaries are smooth. Effects such as reverb or other musical elements may ``leak'' into the succeeding segment, or the succeeding segment may be too different to the current one to feel like its natural continuation. For this reason, we propose an algorithm for identifying smooth transitions around segment boundaries. Given segment $\alpha$ with boundaries $\{p_\alpha, q_\alpha\}$ and a candidate segment $\beta$ with boundaries $\{p_\beta, q_\beta\}$, we search in an area with a radius of 4 measures (or equal to the length of either segment if it is shorter than 4 measures) around $q_\alpha$ and $p_\beta$ for the best entry- and exit-point respectively.

To do this, we use the combined recurrence matrix $\matr{R}$ we used for encoding structure in Section \ref{subsec:structure_encoding}. We can conceptualize the column indices of this matrix as current beats, and the row indices as the target beats, meaning the search area for the transition between $\alpha$ and $\beta$ can be expressed as the square submatrix of $\matr{R}$ defined by columns $q_{\alpha-4}$ to $q_{\alpha+4}$ and rows $p_{\beta-4}$ and $p_{\beta+4}$. We search this area for diagonals of connected beats that would indicate a repeated pattern. Given the number of beats $g$ in each measure of a given music piece, we only consider diagonals whose elements' indices are an integer multiple of $g$ apart, so that we retain the position within the metrical structure during a transition.
 
We select the diagonal that is the longest and the closest to the boundary as long as its length is longer than 1 measure. From this diagonal, we select the midpoint, and store the column index as the entry-point and row index as the exit-point (see Fig. \ref{fig:rec}). We prioritize long patterns and only consider those that are at least a measure long to increase the confidence that a musical idea is repeated and not simply a sole beat being similar to another, a possible issue we found with another system in our evaluation (see Sec. \ref{sec:evaluation}). We use this algorithm for all combinations of non-overlapping current and candidate segments to extract a set of suitable transition points $\mathcal{T}$ anchored to the piece's structural segments.

To prioritize the best transition points in $\mathcal{T}$, as explained further in \ref{subsec:optimization}, we define a metric for the cost $\mathcal{C}(i,j)$ of each transition $\mathcal{T}_{i, j}$. For points on segment boundaries that did not have a more ideal neighbor point transition, we set $\mathcal{C}(i,j)=1$. For other points, we infer the cost from $\matr{R}$, which reflects the feature similarity between two points. If it's a forward transition (i.e. $i>j$), we set $\mathcal{C}(i,j)=1-\matr{R}_{i,j}$, while if it's a backward transitions (i.e. $i<j$), we penalize it such that $\mathcal{C}(i,j)=1-(\matr{R}_{i,j})/4$ so that backwards transitions are discouraged but still an option if the rearrangement aims to extend the music piece.

\subsubsection{Internal transitions points}\label{subsubsec:internal_transition_points}
While spectral clustering allowed us to separate musical ideas and group them by similarity, it is likely that consecutive repetitions will simply be grouped in a single segment. This means that in cases such as a chorus having 4 almost identical repetitions of a chord progression or maybe even lyrics, we may miss the cut-points for skipping some of those repetitions in order to make a short rearrangement. To alleviate this, we traverse some of the middle segmentation levels ($k \in \{4,5,6\}$) that are likely comprised of segments large enough to contain repetitions. On a segment-per-segment basis, we search for diagonals in $\matr{R}$ graphs with the same restrictions used for identifying segment transitions, only this time the search area for a segment $\alpha$ with boundaries $\{p_\alpha, q_\alpha\}$ will be the submatrix defined by columns $p_{\alpha-4}$ to $p_{\alpha+4}$ and rows $p_{\alpha-4}$ and $p_{\alpha+4}$. We add the midpoint of the best diagonal to the set $\mathcal{T}$ of transition points with a cost of $\mathcal{C}(i,j)=1-\matr{R}_{i,j}$.

\subsection{Optimization}\label{subsec:optimization}
Our goal is to produce a music rearrangement of a determined duration, with the additional restriction that we select a first and a last segment from any levels for the start and end of the rearrangement respectively. Although we initially conceptualized rearrangement as the reshuffling of the hierarchical segmentation, we extracted a set of transition points along with their transition cost that we can now use to reframe rearrangement as a path finding problem. We use the approach proposed by Stoller et al.~\cite{loudness} for finding a solution under similar constraints. We want to construct a path defined by a sequence of beats $\matr{A} = \{a_1, a_2,...,a_L\}$ with $a \in B$ where $L$ is the number of beats corresponding to the desired piece length. $\matr{A}$ is constructed by minimizing the cost between two consecutive beats $\mathcal{C}(a_i,a_{i+1})$ subject to: 
\begin{enumerate}
\setlength{\itemsep}{1pt}
\setlength{\parskip}{0pt}
\setlength{\parsep}{0pt}
\item{$a_1=b_1$ and $a_L=b_N$, meaning the rearrangement starts and ends from the same beats as the original;}
\item{a point $a_l=b_i$ only being succeeded by $b_{i+1}$ or $b_j$ if the transition $\mathcal{T}_{i,j}$ exists; and}
\item{the duration of $A$ being within a radius equal to the mean measure duration from the target duration.}
\end{enumerate}
Then, the final cost is computed as the sum over all consecutive beat pairs in $\matr{A}$: $ \sum_{i=1}^{L-1} \mathcal{C}(a_i,a_{i+1})$.

Stoller et al.~\cite{dijkstra} frame this problem as a single-source shortest-paths problem in a directed weighted graph $G=(V, W, a_i)$ with vertex $v_{i,j}\in V$ representing the selection of $b_i$ as $a_j$ and edge $w_{i,j}=(v_{i,k},v_{j,k+1})$, $w_{i,j}\in W$ for every pair of $(a_i,a_j)$. The authors consider all possible transitions between $N$ beats in an $N\times N$ cost matrix. In our case, we only considered transitions that are adequately smooth, with a fallback of transitions during segment boundaries and the possibility of dynamically adjusting the segmentation level. This means that we draw edges only between beats pairs for which a transition exists, including when $b_{i+1}$ succeeds $b_i$, resulting in a much smaller number of possible paths and thus a faster optimization. Similarly to Stoller et al.~\cite{dijkstra}, we use Dijkstra's shortest path algorithm to find the optimal path in $G$.

\section{Evaluation}\label{sec:evaluation}
We conducted a listening study to assess the quality of different music rearrangement systems. Unfortunately, to our best knowledge, no other rearrangement approach is publicly available and open-source. We therefore decided to use the popular Remix tool in Adobe software ~\cite{adoberemix}, a widely-used commercially-available rearrangement tool, through Adobe Audition version 23.0.
We compare it with our own rearrangement system, as well as a manual rearrangement produced by a sound engineer, who was asked to make no further edits apart from deleting, repeating, and rearranging sections and optionally using cross-fades during transitions.

While previous surveys \cite{loudness,cnn,retargeting,dancefloor} focus solely on assessing how perceivable transitions are, we argue for the importance of other musical elements in the rearrangement, such as the structural coherence and lack of repetitions. We therefore provide users with complete rearrangements of 5 songs to rate on a 5-point scale. We use a likert multi-scale experimental design where the participants are asked to rate the songs presented in random order on 4 axes:
(1) \textbf{consistency}: the audio feels consistent and well put together, without any noticeable, distracting, or abrupt discontinuities in rhythm, dynamics, or melody, or other concatenation errors;
(2) \textbf{development}: the structure and musical content are arranged in a sensible order, without abrupt changes in musical ideas, dynamics, or mood;
(3) \textbf{balance}: the piece has a sensible balance of novelty and repetitiveness given its length, without excessive repetitions nor a continuously changing theme;
(4) \textbf{overall quality}: for its duration, the piece can be considered a good, standalone piece of music, with good consistency, development, and balance.

We choose songs from different cultural traditions and styles: 2 popular songs, 1 western classical song, 1 Bollywood song with low recording quality, and 1 Greek island dance song with a noisy phone recording. By varying recording quality we wanted to test the robustness of the automatic methods to audio distortions. We restrict the length of the 5 rearrangements to 38, 42, 46, 50, and 54 seconds to keep the experiment around 20 minutes long. At the start of the experiment, we provide users with short, negative examples for each rating axis. We used the webMUSHRA web listening study interface ~\cite{schoeffler2018webmushra} to conduct the study, presenting the 3 rearrangements of each song (Human, Adobe Remix, and Ours) one after the other in random order, without ever giving any information about the rearrangement systems. A total of 17 participants completed the study.

\begin{figure}[]
    \centering
  \includegraphics[width=0.45\textwidth]{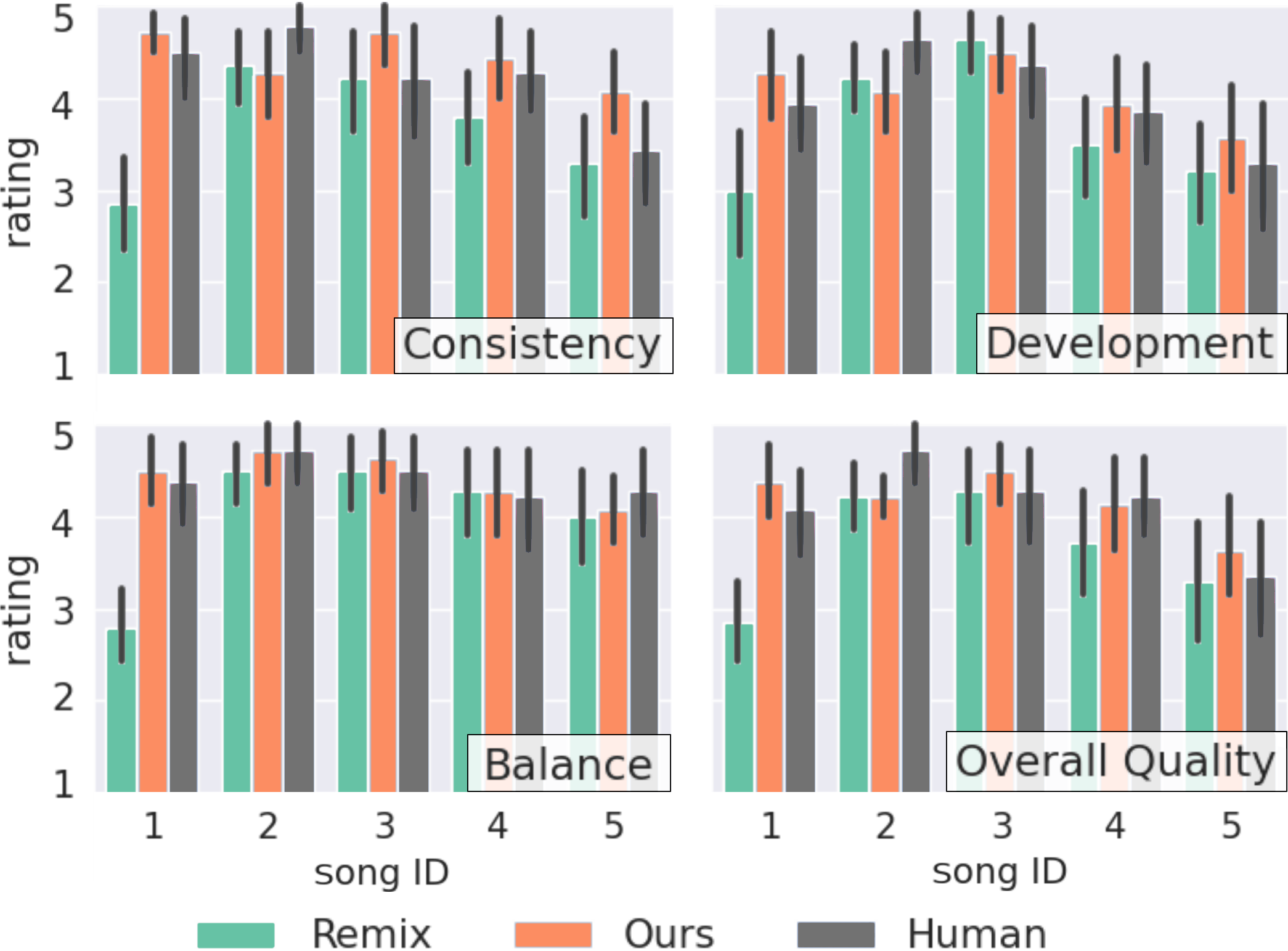}
  \caption{\label{figure}Ratings across 4 axes for each system and song}
  \label{fig:survey}
\end{figure}

The mean ratings for all 5 songs are presented for each system in Fig \ref{fig:survey}. The error bars represent $95\%$ confidence intervals. The manual rearrangement did not receive a perfect rating, although it did collect the highest overall mean rating. Although the sound engineer was satisfied with this rearrangement, manually rearranging a piece requires a lot of subjective judgment, some of which might not be shared by the listeners. This is especially evident in the axis of development, where it is hard to judge what is possible in rearrangements with very small durations. 

Overall, the closed-source Remix tool has lower mean ratings than our open-source system in all axes. A noteable example of a failure is in the case of ``Dancing Queen'' by Abba (ID: 1), where the Remix tool seems to consider some beats with the same chord a viable transition point. This leads to noticeable discontinuities, as the surrounding content does not match, a behavior that inspired the longest diagonal search in our method. Since the error bars are quite large, more participants are required to assess the differences between the three rearrangement versions, and a further listening study should propose more challenging constraints that can stress-test the systems. We note that the Bollywood and the Greek song had lower ratings in terms of quality and this may be explained by the unfamiliarity of the participants with the song, lower audio quality or simply that they are difficult to rearrange. Further experiments are needed to disentangle between preference, recording quality, and rearrangement quality.

\section{Conclusions}\label{sec:conclusions}

In this paper we propose a method for rearranging music recordings that is anchored to their hierarchical structure. We use a semantically and acoustically rich input feature representation to segment pieces and identify smooth transitions points across and within structural segments. 
Experimental results show that on average our structure-oriented approach can produce more consistent musical development, less noticeable cuts, and overall better quality rearrangements, close to what a sound engineer would produce. However, we refrain from generalizing our conclusions because of the limited rearrangements evaluated on a larger scale. Unlike previous approaches that evaluate rearrangement solely on audio snippets that contain transitions~\cite{loudness,cnn,retargeting,dancefloor}, we evaluate axes such as consistency and balance that require the whole rearrangement to be played, therefore increasing the survey time by a lot. Future work includes a larger-scale user evaluation with a larger variety of songs and more qualitative feedback, as well as the investigation of quantitative approaches for evaluating rearrangements on the axes of consistency, development, and balance.
With that said, to our best knowledge our system is the only freely-accessible open-source implementation for music rearrangement, so we encourage readers to experiment with the rearrangement Python package and the listening examples \cite{webpage}.

\vfill\pagebreak

\bibliographystyle{IEEEbib}
\bibliography{strings,refs}

\end{document}